# Dissociation of Liquid Water on Defective Rutile $TiO_2$ (110) Surfaces Using Ab-Initio Molecular Dynamics Simulations

Huili Wang（王会丽）[1,3], Zhenpeng Hu（胡振芃）[1], Hui Li（李晖）[2,‡]

1. School of Physics, Nankai University, Tianjin, 300071, China
2. Beijing Advanced Innovation Center for Soft Matter Science and Engineering, Beijing University of Chemical Technology, Beijing, 100029, China
3. Institute of Physics, Chinese Academy of Sciences, Beijing, 100190, China

Corresponding author. E-mail: [‡]hli@buct.edu.cn (H.L)

In order to obtain a comprehensive understanding of both thermodynamics and kinetics of water dissociation on $TiO_2$, the reactions between liquid water and perfect and defective rutile $TiO_2$ (110) surfaces were investigated using *ab initio* molecular dynamics simulations. The results showed that the free-energy barrier (~ 4.4 kcal/mol) is too high for a spontaneous dissociation of water on the perfect rutile (110) surface at a low temperature. The most stable oxygen vacancy ($Vo_1$) on the rutile (110) surface cannot promote the dissociation of water, while other unstable oxygen vacancies can significantly enhance the water dissociation rate. This is opposite to the general understanding that $Vo_1$ defects are active sites for water dissociation. Furthermore, we reveal that water dissociation is an exothermic reaction, which demonstrates that the dissociated state of the adsorbed water is thermodynamically favorable for both perfect and defective rutile (110) surfaces. The dissociation adsorption of water can also increase the hydrophilicity of $TiO_2$.

# I. Introduction

Since the study by Fujishima *et al*. in 1972 on photocatalysis of titanium dioxide ($TiO_2$) with a photochemical production of hydrogen from water, extensive studies have been performed on photochemistry of $TiO_2$ owing to its applications in various fields, including photochemical water splitting, solar cells, and catalysis [1–4]. Oxygen vacancies are crucial in the reactions on the oxide's surface [5–6]. As the rutile $TiO_2$ (110) surface is the most common and stable surface of this material [7, 8], the dissociation of water on a perfect and defective $TiO_2$ (110) surfaces with different kinds of oxygen vacancy defects has been intensively studied. However, a proper correlation between the experimental and theoretical studies has not been yet established.

According to scanning tunneling microscopy (STM) observations and first-principles calculations, it is generally accepted that at room temperature water can be trapped by bridging oxygen vacancies, accompanied by dissociative adsorption at the bridging oxygen vacancies and formation of two bridging hydroxyls ($OH_b$) [9–15]. Wendt *et al*. reported that the two $OH_b$ are near neighbors and remain stable until they interact with the surrounding water, splitting to single hydroxyl groups [12]. However, several studies argued that the number of dissociative water molecules is larger than the number of bridging oxygen vacancies [16, 17]. For example, Kristoffersen *et al*. revealed a new water-dissociation channel, and showed that oxygen vacancies at the $<1\bar{1}1>$ step edges are active sites for water dissociation contributing with up to two hydrogen adatoms on the terraces [18]. It is not yet known whether other vacancy defects have similar effects on the water dissociation. In addition, there has been a controversy for many years whether water adsorbed at a five-fold coordinated titanium at a perfect $TiO_2$ (110) surface can dissociate [11, 12, 19–22]. Water dissociation attracts an increasing interest as the reaction is exothermal but the large barrier hinders water dissociation at a low temperature [17, 23-25]. Therefore, the *ab initio* molecular dynamics (AIMD) simulations at finite temperature is very necessary for the research of water dissociation on the perfect TiO₂ (110) surface. A possible reason for the contradiction between the different studies of water/TiO₂ (110) interfaces is that most of them considered water in the vapor phase or at a low coverage under ultrahigh-vacuum conditions, which significantly differs from the real situation where TiO₂ interacts with liquid water at ambient conditions. To the best of our knowledge, although many studies have been performed on liquid-water/rutile-(110) interfaces [26–29], there is no systematic study on dynamics of water dissociation at interfaces between liquid water and defective rutile TiO₂ (110) surfaces with different types of oxygen vacancies at ambient conditions.

**Fig. 1.** Ball and stick model of the rutile TiO₂ (110) surface. Red and gray balls represent the O and Ti atoms. The yellow balls represent the different types of oxygen defects.

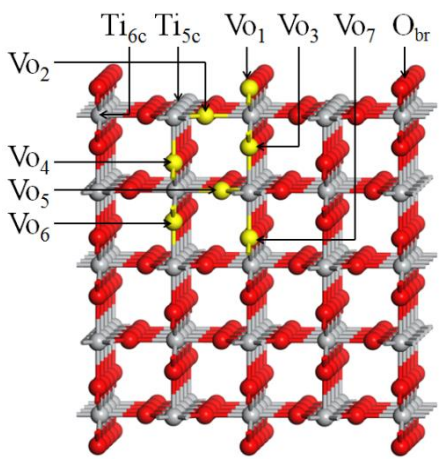

In this study, we systematically investigate the interaction between liquid water and perfect and defective rutile TiO₂ (110) surfaces with different types of oxygen vacancies denoted as $V_{O_1}$, $V_{O_2}$, $V_{O_3}$, and $V_{O_4}$, as shown in Fig. 1. Using AIMD simulations, we reveal the dynamic process of water dissociation at different interfaces in an aqueous environment at room temperature. We show that $V_{O_1}$ and $V_{O_3}$ have higher stabilities, compared with the other defects on the rutile (110) surface; however, their catalytic activities for water dissociation are inferior with respect to those of $V_{O_2}$ and $V_{O_4}$. On the other hand, the 5-coordinated $Ti_{5c}$ site exhibits an unexpectedly higher activity for water dissociation, which indicates that the water dissociation adsorption can occur on the ideal rutile (110) surface, without significant contributions from highly stable oxygen vacancies on the rutile surface. These results provide a detailed understanding of the water dissociation on rutile at the atomic scale in a real environment.

## II. Computational methods

AIMD simulations are performed in the framework of density functional theory (DFT) using the QUICKSTEP code in the CP2K software [30], in order to not only reveal the interatomic interactions determined by the electronic structure, but also investigate the real-time dynamics and thermal fluctuations at a finite temperature. A mixed Gaussian and plane-wave basis set with an energy cutoff of 280 Ry was used for the expansion of the wave functions. The Becke–Lee–Yang–Parr (BLYP) functional was employed for the exchange correlation energy, combined with an empirical correction for the Van der Waals forces [31, 32]. All MD simulations are performed in a constant volume and temperature ensemble with a target temperature of 350 K controlled by a Nose–Hoover thermostat [33]. The simulation time for each MD trajectory is larger than 20 ps with a time step of 1.0 fs. We utilized a five-layer (4 × 2) periodic slab containing three planes with a composition of O–$Ti_2O_2$–O (see Fig. 1) with a vacuum space of 30.0 Å to model the rutile (110) surface. It has been demonstrated that the surface energy oscillates for a weak perturbation of the slab layers, and that a small odd number of slab layers leads to a rapid energy convergence; therefore, the five-layer slab is sufficiently thick for the $TiO_2$ calculations [7, 34, 35]. During the simulations, we fixed the last three layers. The liquid water film in contact with the surface had a thickness of approximately 5 Å, consisting of 32 molecules. The (110) surface consists of rows of five-coordinated Ti atoms ($Ti_{5c}$) and rows of two-coordinated O atoms ($O_{br}$) bridged to six-coordinated Ti atoms ($Ti_{6c}$), as shown in Fig. 1.

## III. Results and discussion

First, we calculate the formation energies of different single oxygen vacancies in vacuum, as shown in Table I. The vacancy formation energy is defined as the energy required to transfer an oxygen atom to the gap phase:

$$E_{vac} = E_{Ti_nO_{2n-1}} + \frac{1}{2}E_{O_2} - E_{Ti_nO_{2n}} \qquad (1)$$

where $E_{Ti_nO_{2n-1}}$ and $E_{Ti_nO_{2n}}$ are the total energies of the defective and perfect slab systems, respectively, and $E_{O_2}$ is the total energy of an isolated $O_2$ molecule. $V_{O1}$ has the lowest formation energy of 3.60 eV, which indicates that this is the most likely vacancy to form. In addition, the $V_{O3}$ and $V_{O7}$ defects form more easily, compared with the $V_{O2}$ and $V_{O4}$ defects. This result is in agreement with that of Oviedo *et al*. [34].

Table I. Formation energies of the different types of oxygen defects shown in Fig. 1.

| Defect | $V_{O1}$ | $V_{O2}$ | $V_{O3}$ | $V_{O4}$ | $V_{O5}$ | $V_{O6}$ | $V_{O7}$ |
|---|---|---|---|---|---|---|---|
| $E_{vac}$/eV | 3.61 | 5.54 | 4.04 | 5.58 | 4.98 | 5.67 | 4.52 |

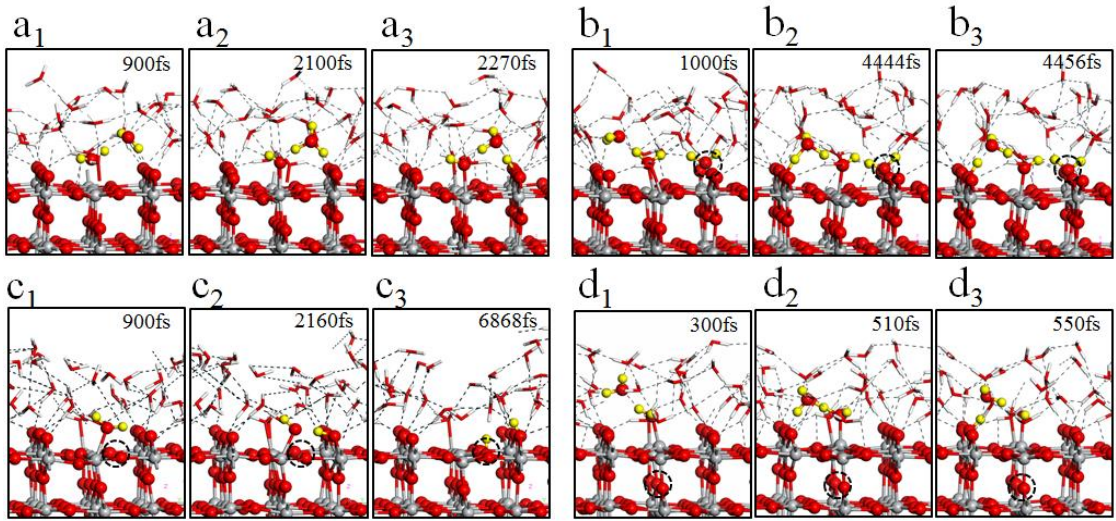

**Fig. 2.** MD snapshots of water dissociation processes on a (a) perfect, (b) $V_{O1}$, (c) $V_{O2}$, and (d) $V_{O4}$ rutile (110) surfaces. The dashed circles outline the locations of the oxygen vacancies.

Using AIMD simulations, a spontaneous water dissociation is observed on a perfect, $V_{O1}$, $V_{O2}$, and $V_{O4}$ rutile (110) surfaces. As shown in Figs. 2(a–d), identical dissociation processes are observed on the perfect and $V_{O1}$ and $V_{O4}$ defective surfaces. The water molecule adsorbed on a $Ti_{5c}$ site gives its proton to another water molecule with a hydrogen bond to the lattice $O_{br}$, forming $H_3O^+$. Another hydrogen-bonded proton of this intermediate structure transfers to the corresponding $O_{br}$, which leads to a formation of $OH_b$ and terminal hydroxyl ($OH_t$). Therefore, our simulations show that intermolecular hydrogen bonds are important in water dissociation, consistent with results in previous studies [26, 27, 36]. At the interface between liquid water and $V_{O1}$ $TiO_2$ (110) surface, a water molecule adsorbed on a $Ti_{5c}$ site dissociates, while a water molecule adsorbed on the $V_{O1}$ site remains unaffected throughout the simulations. This confirms that the $V_{O1}$ defect is not an active site for water dissociation. The water dissociation at the $V_{O2}$ $TiO_2$ (110) surface exhibits a different behavior, as shown in Fig. 2(c). A water molecule adsorbed on $Ti_{5c}$ next to the $V_{O2}$ gives its proton to $O_{br}$ directly through a hydrogen-bonding interaction, and then the terminal hydroxyl populates the $V_{O2}$ defect. In order to provide further insights into the dissociation process, we calculate the distance variation between the bridge oxygen atom and attached proton, and that between the oxygen atom in the dissociative water molecule and its missing proton (Fig. S1, Supporting Information). The results show that the O–H distance becomes larger than 1.0 Å after the water dissociation, and that all bridging hydroxyls can stably exist as the $O_{br}$–H distances remain ~1.0 Å after the water dissociation. A dissociation of only one water molecule is observed during each simulation without any recombination, which indicates that the dissociation of the water molecule can prevent other water molecules to split. It is worth noting that no water dissociation is observed on the $V_{O3}$ defective surface during the simulation, which demonstrates that the $V_{O3}$ can even make the rutile (110) surface more inert to water.

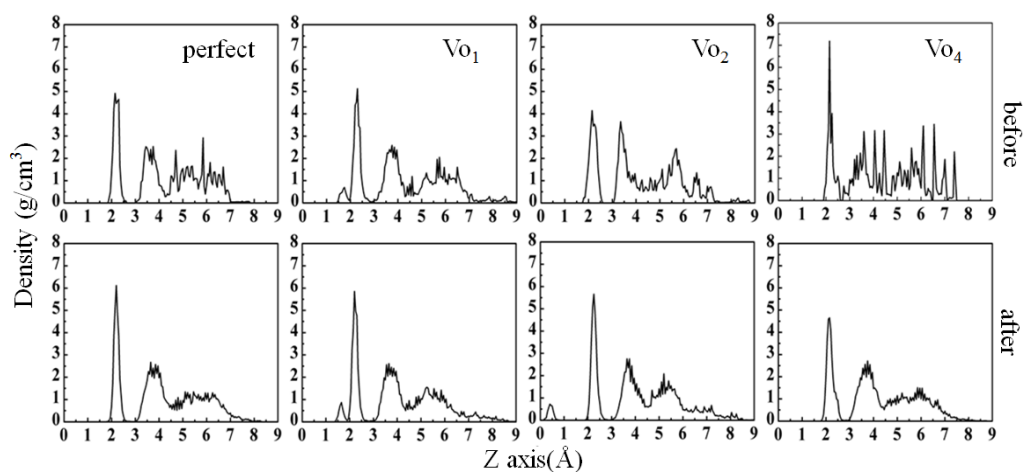

**Fig. 3.** Average densities of water along the surface normal before and after the water dissociation

on a perfect, $Vo_1$, $Vo_2$, and $Vo_4$ $TiO_2$ (110) surfaces.

The average density profiles of water along the surface normal before and after the water dissociation are calculated, as shown in Fig. 3. An obvious stratification of water is observed. Regarding the location of the first (main) adsorption layer, the AIMD predicts a height of ~2.15 Å from the surface, which is in agreement with previous simulation results of 2.10 Å and 2.07 Å [28, 29]. The second layer is centered at a height of ~3.75 Å, consistent with other simulation and experimental results [28, 29]. For the perfect $TiO_2$-(110)/water interface, the first largest peak corresponds to a density of ~4.92 $g/cm^3$ before the water dissociation, smaller than the value of 6.12 $g/cm^3$ after the water dissociation, which indicates that $TiO_2$ is more strongly bonded to water after the water dissociation. A similar phenomenon is observed for the $Vo_1$ and $Vo_2$ defective surfaces; for example, the first peak of the water density increases from 5.13 $g/cm^3$ to 6.08 $g/cm^3$ after the water dissociation for the $Vo_1$ surface, and from 4.14 $g/cm^3$ to 5.66 $g/cm^3$ after the water dissociation for the $Vo_2$ surface. Therefore, the AIMD simulations demonstrate that the surface hydrophilicity of $TiO_2$ is significantly enhanced by the water dissociation. We cannot make a definite conclusion owing to the too fast reaction rate on the $Vo_4$ interface.

Further, we study the thermodynamics of water splitting on the perfect and defective rutile (110) surfaces. According to the energy values before and after the water dissociation listed in Table II, all of these reactions are exothermic, which indicates that the dissociative states are more stable than molecular adsorption states. The dissociation state on the $Vo_2$ interface is the most stable, followed by that on the $Vo_1$ rutile-(110)/water interface. The energy for the water dissociation on the $Vo_4$ defect is not provided, owing to the too fast dissociation reaction, which occurs before the system can be fully equilibrated and reach the target temperature.

**Table II.** Energy changes for the water dissociations at the different surfaces; $\Delta E = E_{aft} - E_{bef}$, where $E_{bef}$ and $E_{aft}$ denote the average energies before and after the water dissociation, respectively.

| interface | perfect | $V_{O1}$ | $V_{O2}$ | $V_{O3}$ |
|---|---|---|---|---|
| $\Delta E$/eV | -0.16 | -1.23 | -2.09 | -0.41 |

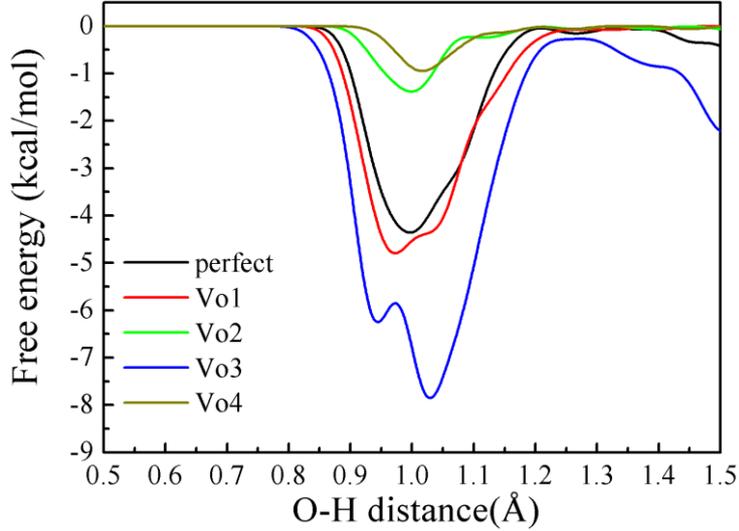

**Fig. 4.** Free energies for water dissociation at the different interfaces.

In order to understand the kinetics of the water dissociation on the different $TiO_2$ surfaces, the free-energy surfaces for the reactions are studied using metadynamics [37, 38] with an appropriate collective variable (CV) as a function of the O–H bond length in the dissociative water. For a complete sampling, numerous Gaussian-type potentials are added to the free-energy surface to assist the system in overcoming a large free-energy barrier. The sum of the Gaussian potentials is opposite to the free-energy; therefore, we can obtain the free-energy of the water dissociation processes, as shown in Fig. 4. During these simulations, one Gaussian hill is spawned every 10 fs; the height of the hill is $10^{-4}$ Ha. The $V_{O2}$ and $V_{O4}$ defects exhibit extremely low free-energy barriers of 1.38 kcal/mol and 1.51 kcal/mol, respectively, which indicates that the water dissociation on these two defects can even occur at a very low temperature. On the other hand, at the perfect and $V_{O1}$ interfaces, the dissociative water molecule adsorbs on the $Ti_{5c}$ sites, so similar free energy barriers for water dissociation on the perfect (4.36 kcal/mol) and $V_{O1}$ (4.79 kcal/mol) rutile (110) surfaces are obtained.

These barriers are high for a reaction that spontaneously occurs at room temperature, which indicates that the dissociation adsorption rates for these two surfaces could be sensitive to temperature, consistent with previous experimental studies [39]. In order to check whether the $V_{O1}$ defect is an active site, we considered five independent MD trajectories. Two trajectories show that the water molecule adsorbed on the $Ti_{5c}$ site dissociates, while that in the $V_{O1}$ site remained unaffected; no water dissociation was observed for the other three trajectories with the $V_{O1}$ defect. In the metadynamics simulation, a free barrier as high as 12.36 kcal/mol is obtained, which indicates that the dissociation of water on the $V_{O1}$ site is kinetically unfavorable. Therefore, according to the MD simulation, the $Ti_{5c}$ site, rather than the $V_{O1}$ site, is the active site for water dissociation. The reaction rate is very sensitive to temperature. The free energy barrier of water dissociation on $V_{O3}$ obtained from the metadynamics is as high as 7.85 kcal/mol, which reveals that the reaction is kinetically unfeasible at ambient conditions. This result also supports the conclusion that $V_{O3}$ is inert to water dissociation.

## IV. Conclusion

Using AIMD simulations, we investigated water dissociation on rutile (110) surfaces. The water molecule adsorbed on the $Ti_{5c}$ site of the perfect rutile (110) surface needed to overcome a free-energy barrier of 4.36 kcal/mol, which was too high for a spontaneous reaction at a low temperature; consequently, no dissociation was observed in some of the low-temperature experiments. In addition, the most common defect ($V_{O1}$) on the rutile (110) surface could not promote the dissociation of water, while other unstable oxygen vacancies such as $V_{O2}$ and $V_{O4}$ could significantly enhance the water dissociation rate. This is opposite to the established understanding that $V_{O1}$ defects are active sites for water dissociation accompanied by two $OH_b$. In addition, the kinetics of water dissociation on the perfect and defective $TiO_2$ surfaces reveal that all dissociation states have higher thermodynamic stabilities than the molecular adsorption state. Furthermore, the dissociation adsorption of water can significantly increase the hydrophilicity of $TiO_2$. Our findings provide a detailed understanding of the interactions between water and rutile surfaces, which is of importance for applications of TiO$_2$ materials.

## Acknowledgement

The authors would like to thank the National Natural Science Fund of China (NSFC) (11374333, 21773005, and 21773124). This study was supported by the Doctoral Fund of the Ministry of Education of China (20120031120033) and Research Program for Advanced and Applied Technology of Tianjin (13JCYBJC36800).